\documentstyle[12pt]{article}

\catcode`\@=11
\long\def\@makefntext#1{ 
\protect\noindent \hbox to 3.2pt {\hskip-.9pt  
$^{{\ninerm\@thefnmark}}$\hfil}#1\hfill} 

\def\thefootnote{\fnsymbol{footnote}}
 \def\@makefnmark{\hbox to 0pt{$^{\@thefnmark}$\hss}}  
	
\def\ps@myheadings{\let\@mkboth\@gobbletwo
\def\@oddhead{\hbox{} 
\rightmark\hfil\ninerm\thepage}   
\def\@oddfoot{}\def\@evenhead{\ninerm\thepage\hfil 
\leftmark\hbox{}}\def\@evenfoot{}
\def\sectionmark##1{}\def\subsectionmark##1{}}

\begin{document}

\newcommand{\symbolfootnote}{\renewcommand{\thefootnote}
	{\fnsymbol{footnote}}}
\renewcommand{\thefootnote}{\fnsymbol{footnote}}
\newcommand{\alphfootnote}
	{\setcounter{footnote}{0}
	 \renewcommand{\thefootnote}{\sevenrm\alph{footnote}}}

\newcounter{sectionc}\newcounter{subsectionc}\newcounter{subsubsectionc}
\renewcommand{\section}[1] {\vspace{0.6cm}\addtocounter{sectionc}{1} 
\setcounter{subsectionc}{0}\setcounter{subsubsectionc}{0}\noindent 
	{\bf\thesectionc. #1}\par\vspace{0.4cm}}
\renewcommand{\subsection}[1] {\vspace{0.6cm}\addtocounter{subsectionc}{1} 
	\setcounter{subsubsectionc}{0}\noindent 
	{\it\thesectionc.\thesubsectionc. #1}\par\vspace{0.4cm}}
\renewcommand{\subsubsection}[1] {\vspace{0.6cm}\addtocounter{subsubsectionc}{1}
	\noindent {\rm\thesectionc.\thesubsectionc.\thesubsubsectionc. 
	#1}\par\vspace{0.4cm}}
\newcommand{\nonumsection}[1] {\vspace{0.6cm}\noindent{\bf #1}
	\par\vspace{0.4cm}}
					         
\newcounter{appendixc}
\newcounter{subappendixc}[appendixc]
\newcounter{subsubappendixc}[subappendixc]
\renewcommand{\thesubappendixc}{\Alph{appendixc}.\arabic{subappendixc}}
\renewcommand{\thesubsubappendixc}
	{\Alph{appendixc}.\arabic{subappendixc}.\arabic{subsubappendixc}}

\renewcommand{\appendix}[1] {\vspace{0.6cm}
        \refstepcounter{appendixc}
        \setcounter{figure}{0}
        \setcounter{table}{0}
        \setcounter{equation}{0}
        \renewcommand{\thefigure}{\Alph{appendixc}.\arabic{figure}}
        \renewcommand{\thetable}{\Alph{appendixc}.\arabic{table}}
        \renewcommand{\theappendixc}{\Alph{appendixc}}
        \renewcommand{\theequation}{\Alph{appendixc}.\arabic{equation}}
        \noindent{\bf Appendix \theappendixc #1}\par\vspace{0.4cm}}
\newcommand{\subappendix}[1] {\vspace{0.6cm}
        \refstepcounter{subappendixc}
        \noindent{\bf Appendix \thesubappendixc. #1}\par\vspace{0.4cm}}
\newcommand{\subsubappendix}[1] {\vspace{0.6cm}
        \refstepcounter{subsubappendixc}
        \noindent{\it Appendix \thesubsubappendixc. #1}
	\par\vspace{0.4cm}}

\def\abstracts#1{{
	\centering{\begin{minipage}{30pc}\tenrm\baselineskip=12pt\noindent
	\centerline{\tenrm ABSTRACT}\vspace{0.3cm}
	\parindent=0pt #1
	\end{minipage} }\par}} 

\newcommand{\bibit}{\it}
\newcommand{\bibbf}{\bf}
\renewenvironment{thebibliography}[1]
	{\begin{list}{\arabic{enumi}.}
	{\usecounter{enumi}\setlength{\parsep}{0pt}
\setlength{\leftmargin 1.25cm}{\rightmargin 0pt}
	 \setlength{\itemsep}{0pt} \settowidth
	{\labelwidth}{#1.}\sloppy}}{\end{list}}

\topsep=0in\parsep=0in\itemsep=0in
\parindent=1.5pc

\newcounter{itemlistc}
\newcounter{romanlistc}
\newcounter{alphlistc}
\newcounter{arabiclistc}
\newenvironment{itemlist}
    	{\setcounter{itemlistc}{0}
	 \begin{list}{$\bullet$}
	{\usecounter{itemlistc}
	 \setlength{\parsep}{0pt}
	 \setlength{\itemsep}{0pt}}}{\end{list}}

\newenvironment{romanlist}
	{\setcounter{romanlistc}{0}
	 \begin{list}{$($\roman{romanlistc}$)$}
	{\usecounter{romanlistc}
	 \setlength{\parsep}{0pt}
	 \setlength{\itemsep}{0pt}}}{\end{list}}

\newenvironment{alphlist}
	{\setcounter{alphlistc}{0}
	 \begin{list}{$($\alph{alphlistc}$)$}
	{\usecounter{alphlistc}
	 \setlength{\parsep}{0pt}
	 \setlength{\itemsep}{0pt}}}{\end{list}}

\newenvironment{arabiclist}
	{\setcounter{arabiclistc}{0}
	 \begin{list}{\arabic{arabiclistc}}
	{\usecounter{arabiclistc}
	 \setlength{\parsep}{0pt}
	 \setlength{\itemsep}{0pt}}}{\end{list}}

\newcommand{\fcaption}[1]{
        \refstepcounter{figure}
        \setbox\@tempboxa = \hbox{\tenrm Fig.~\thefigure. #1}
        \ifdim \wd\@tempboxa > 6in
           {\begin{center}
        \parbox{6in}{\tenrm\baselineskip=12pt Fig.~\thefigure. #1 }
            \end{center}}
        \else
             {\begin{center}
             {\tenrm Fig.~\thefigure. #1}
              \end{center}}
        \fi}

\newcommand{\tcaption}[1]{
        \refstepcounter{table}
        \setbox\@tempboxa = \hbox{\tenrm Table~\thetable. #1}
        \ifdim \wd\@tempboxa > 6in
           {\begin{center}
        \parbox{6in}{\tenrm\baselineskip=12pt Table~\thetable. #1 }
            \end{center}}
        \else
             {\begin{center}
             {\tenrm Table~\thetable. #1}
              \end{center}}
        \fi}

\def\@citex[#1]#2{\if@filesw\immediate\write\@auxout
	{\string\citation{#2}}\fi
\def\@citea{}\@cite{\@for\@citeb:=#2\do
	{\@citea\def\@citea{,}\@ifundefined
	{b@\@citeb}{{\bf ?}\@warning
	{Citation `\@citeb' on page \thepage \space undefined}}
	{\csname b@\@citeb\endcsname}}}{#1}}

\newif\if@cghi
\def\cite{\@cghitrue\@ifnextchar [{\@tempswatrue
	\@citex}{\@tempswafalse\@citex[]}}
\def\citelow{\@cghifalse\@ifnextchar [{\@tempswatrue
	\@citex}{\@tempswafalse\@citex[]}}
\def\@cite#1#2{{$\null^{#1}$\if@tempswa\typeout
	{IJCGA warning: optional citation argument 
	ignored: `#2'} \fi}}
\newcommand{\citeup}{\cite}

\def\fnm#1{$^{\mbox{\scriptsize #1}}$}
\def\fnt#1#2{\footnotetext{\kern-.3em
	{$^{\mbox{\sevenrm #1}}$}{#2}}}

\font\twelvebf=cmbx10 scaled\magstep 1
\font\twelverm=cmr10 scaled\magstep 1
\font\twelveit=cmti10 scaled\magstep 1
\font\elevenbfit=cmbxti10 scaled\magstephalf
\font\elevenbf=cmbx10 scaled\magstephalf
\font\elevenrm=cmr10 scaled\magstephalf
\font\elevenit=cmti10 scaled\magstephalf
\font\bfit=cmbxti10
\font\tenbf=cmbx10
\font\tenrm=cmr10
\font\tenit=cmti10
\font\ninebf=cmbx9
\font\ninerm=cmr9
\font\nineit=cmti9
\font\eightbf=cmbx8
\font\eightrm=cmr8
\font\eightit=cmti8



\def\beqa{\begin{eqnarray}}
\def\eeqa{\end{eqnarray}}
\def\beq{\begin{equation}}
\def\eeq{\end{equation}}
\def\lab{\label}

\def\ad{\dot{a}}
\def\vol{\int d^4x\,\sqrt{-g}} 
\def\grav{\frac{1}{16 \pi G}}
\def\half{\frac{1}{2}}
\def\gu{g^{\mu\nu}}
\def\gd{g_{\mu\nu}}
\def\diag{{\rm ~diag}}
\def\sign{{\rm ~sign}}
\def\ln{{\rm ~ln}}
\def\dro{\delta\rho/\rho}
\def\hm{h^{-1}{\rm~Mpc}}
\def\vr{\mbox{\bf r}}
\def\vk{\mbox{\bf k}}
\def\vx{\mbox{\bf x}}
\def\vy{\mbox{\bf y}}
\def\cz{\chi_0}

\def\umu{^{\mu}}
\def\unu{^{\nu}}
\def\dmu{_{\mu}}  
\def\dnu{_{\nu}}
\def\umunu{^{\mu\nu}}
\def\dmunu{_{\mu\nu}}
\def\ua{^{\alpha}}  
\def\ub{^{\beta}}
\def\da{_{\alpha}}
\def\db{_{\beta}}
\def\ug{^{\gamma}}
\def\dg{_{\gamma}}
\def\ur{^{\rho}}
\def\dr{_{\rho}}
\def\ut{^{\tau}}
\def\dt{_{\tau}}
\def\uamu{^{\alpha\mu}}
\def\uanu{^{\alpha\nu}}
\def\uab{^{\alpha\beta}}
\def\dab{_{\alpha\beta}}
\def\dabgd{_{\alpha\beta\gamma\delta}}
\def\uabgd{^{\alpha\beta\gamma\delta}}
\def\udeab{^{;\alpha\beta}}
\def\ddeab{_{;\alpha\beta}}
\def\ddemunu{_{;\mu\nu}}
\def\udemunu{^{;\mu\nu}}
\def\dmunurota{_{\mu\nu\rho\tau}}
\def\umunurota{^{\mu\nu\rho\tau}}
\def\ddes{_{;\sigma}}\def\udes{^{;\sigma}}
\def\ddemu{_{;\mu}}  \def\udemu{^{;\mu}}
\def\ddenu{_{;\nu}}  \def\udenu{^{;\nu}}
\def\ddea{_{;\alpha}}  \def\udea{^{;\alpha}}
\def\ddeb{_{;\beta}}  \def\udeb{^{;\beta}}

\def\naba{\nabla_{\alpha}}
\def\nabb{\nabla_{\beta}}
\def\pmu{\partial_{\mu}}
\def\pnu{\partial_{\nu}}
\def\p{\partial}

\let\lam=\lambda  \let\Lam=\Lambda
\let\eps=\varepsilon
\let\gam=\gamma
\let\Gam=\Gamma
\let\alp=\alpha
\let\sig=\sigma
\let\ome=\omega

\def\etal{{\it et al.}\ }
\def\ie{{\it i.e. }\ }
\def\eg{{\it e.g. }\ }

\def\rjmp{ J. Math. Phys. }
\def\rpr{ Phys. Rev. }
\def\rprd{ Phys. Rev. \ {\bf D}}
\def\rprl{ Phys. Rev. Lett. }
\def\rpl{ Phys. Lett. }
\def\rnp{ Nucl. Phys. }
\def\rmodpl{ Mod. Phys. Lett. }
\def\rijmp{ Int. J. Mod. Phys. }
\def\rcmp{ Commun. Math. Phys. }
\def\rcqg{ Class. Quantum Gravit. }
\def\rap{ Ann. Phys. (N.Y.) }
\def\rspj{ Sov. Phys. JETP }
\def\rspjl{ Sov. Phys. JETP Lett. }
\def\rprs{ Proc. R. Soc. }
\def\rgrg{ Gen. Relativ. Gravit. }
\def\rnat{ Nature }
\def\rapj{ Ap. J. }
\def\rapjl{ Ap. J. Lett. }
\def\raaa{ Astron. Astrophys. }
\def\rncim{ Nuovo Cimento }
\def\rptp{ Prog. Theor. Phys. }
\def\raip{ Adv. Phys. }
\def\rjpamg{ J. Phys. A: Math. Gen. }
\def\rmnras{ Mon. Not. R. Ast. Soc. }
\def\rprep{ Phys. Rep. }
\def\rncb{ Il Nuovo Cimento ``B'' }
\def\rssr{ Space Sci. Rev. }
\def\rpasp{ Pub. A. S. P. }
\def\raraa{ Ann. Rev. Astr. Ap. }
\def\rasr{ Adv. Space Res. }

\thispagestyle{\empty}
\baselineskip=16pt
\centerline{\tenbf NON-GAUSSIAN LIKELIHOOD FUNCTION
\footnote{Published in Astro. Lett. and Communications, 1996,
33, 63}}
\vspace{0.8cm}
\centerline{\tenrm LUCA AMENDOLA}
\baselineskip=13pt
\centerline{\tenit Osservatorio Astronomico di Roma}
\baselineskip=12pt
\centerline{\tenit Viale del Parco Mellini, 84,  Rome 00136 - Italy}
\vspace{0.3cm}
\vspace{0.9cm}
\abstracts{
I present here a generalization of the
maximum likelihood method and the $\chi^2$ method
to the cases in which the data are
{\it not} assumed to be Gaussian distributed.
The method, based on the
multivariate Edgeworth expansion,
 can find several astrophysical
applications. I mention only two of them.
First,  in the microwave background analysis, where it cannot
be excluded that the initial perturbations are non-Gaussian.
Second, in the large scale structure statistics, as we already know that the
galaxy distribution
deviates from Gaussianity on the scales
at which non-linearity is important.
As a first interesting result I show here how the confidence regions 
are modified when non-Gaussianity
is taken into account.
}

\vfil
\rm\baselineskip=14pt
\section{Introduction}
\label{sec:intro}

The role of statistics in large scale astrophysics is increasing at a very
fast rate, barely keeping the pace with the flow of observational data
from galaxy surveys and microwave background.
Since we want not only to describe our Universe but also to understand it,
we need quantitative ways to compare observations with theoretical
models. This requires the choice of good statistical descriptors,
like correlation functions, higher-order
moments and similar, {\it and} the ability to determine their
confidence regions (CR), i.e. the probability density function
of the estimators.
The general problem is that, while we certainly need some basic
assumptions with respect to the statistical nature of the data,
we want to keep these assumptions to a minimum. For instance, we would
like to analyze the signal from the cosmic microwave background
(CMB)  experiments or
from galaxy surveys without
assuming that the data are Gaussian distributed.

To this scope I present here a general method to approach such
problems, based on the Multivariate
Edgeworth Expansion (MEE),  an Hermite expansion around a Gaussian
distribution\cite{C67}$^,$\cite{M84}.
 This method is suitable to
the cases in which the data can be reasonably assumed to be
mildly non-Gaussian, and we wish to estimate the region of
confidence of the relevant parameters without the Gaussian assumption.
I can think of several applications of the MEE to the
analysis of astrophysical data. In the
case of the CMB, we can use the MEE to estimate at the
same time such fundamental parameters like the primordial slope
$n$ and the
quadrupole amplitude $Q_{rms}^{PS}$, and
 higher-order parameters like the skewness, 
 along with their confidence regions.
 I will show that
the CR may broaden or contract with respect to the Gaussian case.
Similarly, one can analyse in the same manner the large scale 
structure of galaxy
clustering. Another application is to the case of $\chi^2$ fitting
when the data are not Gaussian.
More details on the method and on its applications 
can be found in another work\cite{A94b}.

\section{Formalism}
Let $d^i$ be a set of experimental data (e.g.,
CMB fluctuations, or galaxy counts), $i=1,..N$, 
and let us form the variables $x^i=d^i-t^i$, where $t^i$ are
the theoretically expected values for the measured quantities.
Let $c^{ij}$ be  the correlation matrix
\beq\label{cm}
c^{ij}=<x^ix^j>\,,
\eeq
and let us introduce the higher-order cumulant matrices
(or  correlation functions)
\beqa
k^{ijk}&=&<x^i x^j x^k>\qquad{\rm~(skewness~matrix)},\\
k^{ijkl}&=&<x^i x^j x^k x^l>-c^{ij}c^{kl}-c^{ik}c^{jl}-c^{il}c^{jk}
\qquad {\rm~(kurtosis~matrix)}\,.
\eeqa
The correlation matrices depend in general both on a
number of theoretical parameters $\alpha_j$, $j=1,..P$
 and on the experimental errors.
We assume the latter to be
Gaussian distributed and
completely characterized by the correlation matrix $e^{ij}$,
 to be
added in quadrature to  the 2-point
correlation function. 
It is useful to define then the matrix
$\lam_{ij}=(c^{ij}+e^{ij})^{-1}\,.$
The parameters $\alpha_j$
are fixed by maximizing, with respect to the
parameters, the likelihood
function 
$~L=f(\vx)\,,$
where $f(\vx)$ is the  multivariate probability
distribution function (PDF) for
 the random variables $x_i$. 
 The usual 
simplifying assumption is then
that $f(\vx)$ is a multivariate Gaussian distribution
\beq
L_g=f(\vx)=G(\vx,\lam)\equiv 
(2\pi)^{-N/2} |\lam|^{1/2} \exp(- x^i \lam_{ij} x^j/2)\,.
\eeq
where $|\lam|={\rm det}(\lam_{ij})$.
A straightforward way to 
generalize the LF so as to include the higher-order correlation functions,
which embody the non-Gaussian properties of the data,
is provided by the 
MEE.
An unknown PDF $f(\vx)$ can indeed be expanded around a multivariate
Gaussian $G(x,\lam)$  according to the formula
\cite{C67}$^,$\cite{M84}$^,$\cite{KSO87}
\beq\label{mee}
f(\vx)=G(\vx,\lam)[1+{1\over 6}k^{ijk}h_{ijk}(\vx,\lam)
+{1\over 24 }k^{ijkl}h_{ijkl}(\vx,\lam)
+{1\over 72 }k^{ijk}k^{lmn}h_{i..n}(\vx,\lam)+...]\,,
\eeq
where $h_{ij..}$ are Hermite tensors,  
a generalization of the Hermite polynomials. If there are $r$ subscripts,
the Hermite tensor $h_{ij..}$ is said to be of order $r$, and is given by
\beq\label{defpol}
h_{ij...}=(-1)^r G^{-1}(\vx,\lam) \partial_{ij...} G(\vx,\lam)\,,
\eeq
where $\partial_{ij...}=(\partial/\partial x_i)(\partial/\partial x_j)...$.
It can be shown that the MEE gives a good approximation to any
distribution function provided that  the cumulants obey the same
order-of-magnitude scaling of a standardized mean.\cite{C67}
This condition is satisfied, for instance, by the cumulants
of the galaxy clustering in the scaling regime, which explains
why the (univariate)
Edgeworth expansion well approximates the probability
distribution of the large scale density field\cite{J93}$^,$
\cite{KB94}
In the past years, the MEE has been employed also to approximate the
biased density distribution for large value of the biasing threshold,
to the scope of calculating the peak correlation functions for
non-G random fields\cite{MLB86} 
and other
descriptors of excursion sets\cite{CLM88}.
The same expansion has been also applied to the statistics of
pencil-beam surveys\cite{A94a},
and to go beyond the Gaussian approximation
in calculating the
topological genus of weakly non-Gaussian fields\cite{M94}.
Let us also note that the MEE can also be immediately
generalized to the case of experimental errors  {\it not} Gaussian distributed.

\section{Best estimators}
The best parameter estimates  are  obtained by
maximizing Eq. (\ref{mee}) with respect to the parameters.
To illustrate some interesting points, let us put ourselves in the
 simplest case, in which all data are independent, and we only
need to estimate the parameters $\sigma$ and $k_3$ defined as:
$
c_{ij}=\sigma^2 \delta_{ij},\,\,
k_{ijk}=k_3 \delta_{ij}\delta_{jk}\,.
$
The maximum likelihood estimators for the variance and the
skewness are then obtained by putting
\beq\label{bestsig}
{d\log L\over d\sigma}=0\,,\qquad
{d\log L\over dk_3}=0\,.
\eeq
The solution  reduce then to
the usual sample quantities
\beq\label{mlvar}
\hat\sigma^2=\sum_i x_i^2/N\,,\qquad
\label{sample}
\hat k_3=\sum_i x_i^3/N\,.
\eeq
(which are asymptotically unbiassed).
The same calculation can be carried out in the more general case
of dipendent variables, but the search for the maximum is more
simply performed numerically when the situation is more complicated. 

Once we have the best estimators $\hat \alpha_i(\vx)$
of our parameters, we need to estimate
the confidence regions for that paramaters. 
The problem consists in determining  the behavior of the
unknown distribution $P[\hat\alpha_i(\vx)]$, when we know the
distribution for the random variables $x_i$. This problem is
generally unsoluble analitycally, and the common approach is
to resort to MonteCarlo simulations of the data. However,
we can always approximate $P(\hat\alpha_i)$ {\it around its peak}
by a Gaussian distribution multivariate {\it in the parameter space};
if the number of data $N\to\infty$, this procedure can be justified
by the central limit theorem. 
 The covariance matrix of the parameters is then\cite{KSO87}
\beq
\Sigma_{ab}^{-1}=-{\partial \log L(\vx,\alpha_a)\over
\partial \alpha_a\partial\alpha_b}\Big|_{\alpha_a=\hat\alpha_a}\,,
\eeq
where $a,b$ run over the dimensionality $P$ of the parameter space.
The component $\Sigma_{22}$, 
i.e. the variance of $\hat k_3$,   is then simply
(dropping the hats here and below)
\beq
\Sigma_{22}= 6\sigma^6/N\,,
\eeq
which, not unexpectedly, is the sample skewness variance, i.e.
the scatter in the skewness of Gaussian samples.
 More interesting is the error in the variance parameter $\sigma$ when
not only a non-zero skewness $k_3$ is present, but also a non-zero
kurtosis parameter $k_4$, defined in a way similar to $k_3$ as
$k_{ijkl}=k_4 s_{ijkl}$. The result is
\beq\label{varvar}
\Sigma_{11}=(\sigma^2/ 2N)\left[
1+\gam_2/2\right]\,,
\eeq
where $\gam_2=k_4/ \sigma^4$ is the dimensionless kurtosis.
Notice that,
in the mild non-Gaussianity condition we are assuming
throughout this work,
 the mixed components  $\Sigma_{12}^{-1}=\Sigma_{21}^{-1}$
are negligible.
The first term in (\ref{varvar})
is the usual variance of the sample variance for
Gaussian, independent data.
 The second term is due to the
kurtosis correction: it will broaden the CR for $\sigma$ when $k_4$
 is positive,
and will shrink it when it is negative.
Depending on the relative amplitude of the higher-order
corrections, the CR for the variance can extend or reduce.
It is important however to remark that this estimate of the
confidence regions is approximated, and that it can be trusted only
around the peak of the likelihood function. 

\section{Non-Gaussian $\chi^2$ method}

If our data are distributed following the MEE, then we can measure
the likelihood to have found our actual dataset integrating
the LF over all the possible outcomes of our experiment.
Then the relevant integral we have to deal with is
\beq\label{rel}
M(\chi_0)=\int_{\chi^2\le\chi^2_0} L(x,\lam) \prod_i dx_i\,,
\eeq
where the region of integration extends over all the possible
data values which lie inside the region delimited by 
the actual  value $\chi^2_0$. 
We can then use $M(\chi_0)$ for evaluating a CR for the
parameters which enter $\chi_0^2$, like the quadrupole and the primordial
slope in the case of CMB. The CR will depend parametrically
on the higher-order moments; however, this will not provide a CR
for the higher-order moments themselves.  The method of the
previous section can always be employed to yield a first
approximation for such moments. 
Fixing a confidence level of
$1-\eps$, we will consider as acceptables the values of the parameters for
which $M(\chi_0)$ is larger than $\eps/2$ and smaller than $1-\eps/2$.
The evaluation of
 (\ref{rel}) would require some discussion\cite{A94b}. Here,
 however,
   I only state the final result:
\beqa\label{finres}
&&M(\cz)=\int L \prod dx_i=F_N(\cz)\nonumber\\
&&+{G_N(\chi_0)\pi^{N/2}\cz^N \over 2\Gamma(2+N/2)}
\left[ C_a\left(N+2-\chi_0^2\right)+C_b
\left(-N-2+2\chi_0^2-{\chi_0^4\over N+4}\right)\right]\,,
\eeqa
where $F_N(\cz)$ is the usual $\chi^2$ cumulative function, and
$C_a=c_1+3 c_2$, and $C_b=c_3+3 c_4+15 c_5$, and
the coefficients $c_i$ are formed\cite{A94b} by summing over all
the even diagonals of the correlation tensors $k^{ij..}$ and multiplying
for the Edgeworth coefficients $(1/24)$ for $c_1,c_2$ and $(1/72)$
for $c_3,c_4$ and $c_5$.
Let us make some comments on Eq. (\ref{finres}). 
First, the fact that
$M(\cz)$ is a cumulative function  provides a simple way to check
the consistency of our assumptions: when the higher-order moments
are too large, the MEE breaks down,  $M(\cz)$ is no longer
monotonic, and can decrease below zero or above unity.
Second, let us suppose that
the higher-order correlation
functions are positive, which is the case for the galaxy
clustering.
Then the non-G corrections in Eq. (\ref{finres})
are negative for  $\cz^2\gg N$. 
The fact that the corrections
are negative for $\cz^2\gg N$  implies that the value of
$\cz=\cz(\eps)$   is larger than in the purely
Gaussian case, in the limit of $\eps\to 0$. 
 Consequently, if the higher-order
correlation functions are positive,
{\it the confidence regions are
sistematically widened when the non-Gaussian corrections
are taken into account}. 
Finally, it is easy to write down
the result in the particular case in which all the cumulant
matrices are diagonal, i.e. for statistically independent variables.
In this case the variables $y^i$ are simply equal to $x^i/\sigma_i$, if 
$\sigma_i=(\lam^i_i)^{-1/2}$, and we can put
 $k^{iii}(y)=k^{iii}(x)/\sigma^3\equiv\gam_{1,i}$,
  and likewise $k^{iiii}(y)\equiv \gam_{2,i}$
(skewness and kurtosis coefficients). Then, we have $c_1=c_3=c_4=0$, and
Eq. (\ref{finres}) can be simplified to 
\beq\label{resind}
M(\cz)=F_N(\cz)+G_N(\chi_0) q(\cz)\,,
\eeq
where
\beq\label{indp}
q(\cz)={6\pi^{N/2}\chi_0^N \over (N+2)\Gamma(N/2)}
\left\{
{\gam_2\over 24}\left[(N+2)-\chi_0^2\right]+{5\over 72}\gam_1^2
\left[-(N+2)+2\chi_0^2-{\chi_0^4\over N+4}\right]\right\}\,,
\eeq
and where  the average squared skewness, $\gam_1^2=
\sum \gam_{1,i}^2/ N,$ and the average kurtosis,
$\gam_2=\sum \gam_{2,i}/N$, have been introduced.

Let me now illustrate graphically 
 some properties of the
 function $M(\cz)$ in its
simplified version (\ref{resind}) above.
In all this section we can think of
 $\cz$ as depending monotonically on
one single parameter, for instance the overall normalization $A>0$
of the correlation function: 
$\cz^2(A)=x^i x^j (A c_{ij}+e_{ij})^{-1}$.
We can then speak of a CR on $\cz$ meaning in fact the corresponding
CR on the parameter $A$. In the general case, the relation
between $\cz$ and its parameters can be quite more complicated.
In Fig. 1{\it a} (for $\gamma_1=0$ and $N=10$),
 I show 
how the function $M(\cz)$ varies with respect to $\gamma_2$.
Schematically, for $\cz^2/N> 1$, the function $M(\cz)$ decreases
when $\gamma_2>0$ and increases in the opposite case. 
As anticipated,
for too large a $\gam_2$, $M(\cz)$ develops a non-monotonic behavior.
The consequence of the behavior of $M(\cz)$
 on the confidence region of
$\cz$ is represented in Fig. 1{\it b}, where the contour plots of the
surface $M(\cz,\gamma_2)$ are shown. 
Consider for instance the two outer contours, corresponding to
$M=.01$, the leftmost, and $M=.99$, the rightmost.
The range of $\cz$ inside
such confidence levels increases for increasing $\gamma_2$.
The same is true for the other contour levels, although
with a  less remarkable trend. This behavior confirms the approximate result
of Eq. (\ref{varvar}).
As anticipated,  this means
that the non-G confidence regions  will be larger and larger 
(if the higher moments are positive) than
the corresponding Gaussian regions for higher and higher 
probability thresholds. 

\begin{figure}
\vspace{3.4in}
\caption{
{\it a)} Plot of $M(\cz)$ as a function of $\cz^2/N$ and of the
dimensionless kurtosis $\gam_2$, for $\gam_1=0, N=10$.
 For $\gam_2=0$ we return to the usual $\chi^2$
cumulative function. 
{\it b)} Contour levels of $M(\cz)$ corresponding to
$M=.01,.1,.2,.3,.7,.8,.9,.01$, from left to right. Notice how
the limits for $\cz$ broaden for increasing $\gam_2$.}
\end{figure}
\begin{figure}
\vspace{3.4in}
\caption{
{\it a)} 
Same as in Fig. 1{\it a}, now with  $\gam_2=0, ~N=100$,
and varying $\gam_1$.
{\it b)}  Contour levels of $M(\cz)$ for the same values as
 in Fig. 1{\it b}. }
\end{figure}

The situation is qualitatively different considering $\gamma_2=0$
and varying $\gam_1$, the average skewness (Fig. 2, with
$N=100$).   
For the outer contours, 
delimiting levels of 1\% on both tails, the CR of $\cz$ {\it increases}
for larger $|\gam_1|$, 
with a minimum for the Gaussian case.
For the internal contours, however, the CR actually shrinks for
larger $|\gam_1|$,   being maximal at the Gaussian point. 
It is clear that in the general case, $\gam_1,\gam_2\not=0$,
the topography of the LF can be quite complicated.

\section{Conclusions}

Let us  summarize the results reported here.
This work is aimed at presenting a new analytic formalism for parametric
estimation with the maximum likelihood method for non-Gaussian
random fields. The method can be applied to a large class
of astrophysical problems. 
The non-Gaussian likelihood function allows the determination
of a full set of parameters and their {\it joint} confidence region,
without arbitrarily fixing some of them, 
 as long as enough non-linear terms are
included in the expansion. 
The CR for all the relevant parameters can be estimated
by approximating the distribution function for the parameter
estimators around its peak by a Gaussian, as in Sect. 3.
 To overcome this
level of approximation, in Sect. 4
I  generalized the $\chi^2$ method to
include non-Gaussian corrections.
The most interesting result is then
that the CR
for the parameters which enter $\cz^2$ is systematically
widened by the inclusion
of the non-Gaussian terms, in the limit of $\eps\to 0$. Two
experiments producing incompatible results can then be brought
to agreement when third and fourth-order cumulants are introduced.

There are two main limitations to the method. One is that one
obviously has to truncate the MEE to some  order, and consequently
 the data analysis  implicitly assumes  that all
the  higher moments   vanish.
The second limitation  is that the method is not applicable to strongly
non-Gaussian field, where the MEE breaks down. 
This can be seen directly from Eq. (\ref{finres}): for arbitrarily
large constants $c_1-c_5$ the likelihood integral is not positive-definite,
although always converge to unity.

\vspace{.2in}
\centerline{\bf Acknowledgments}
I thank St\'ephane Colombi, Scott Dodelson, Sabino Matarrese and
Albert Stebbins for several comments and suggestions at various stages
of this work.

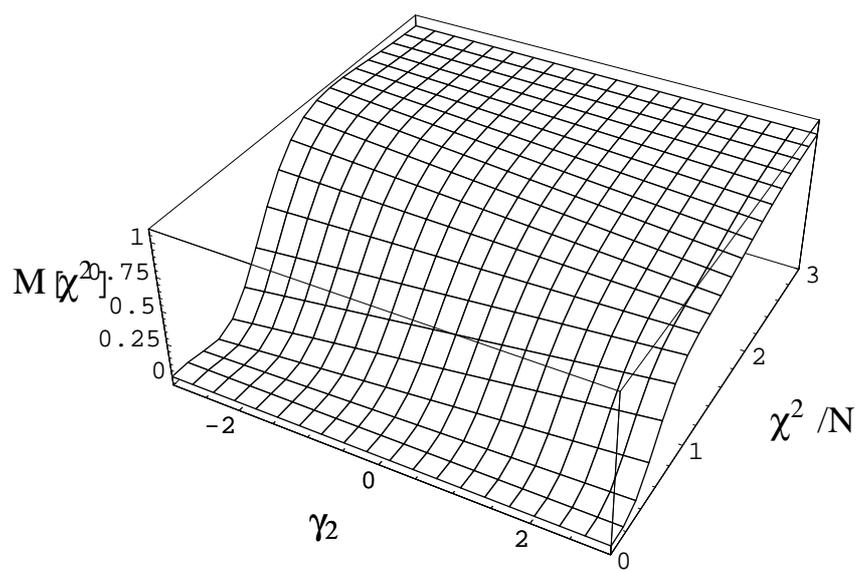
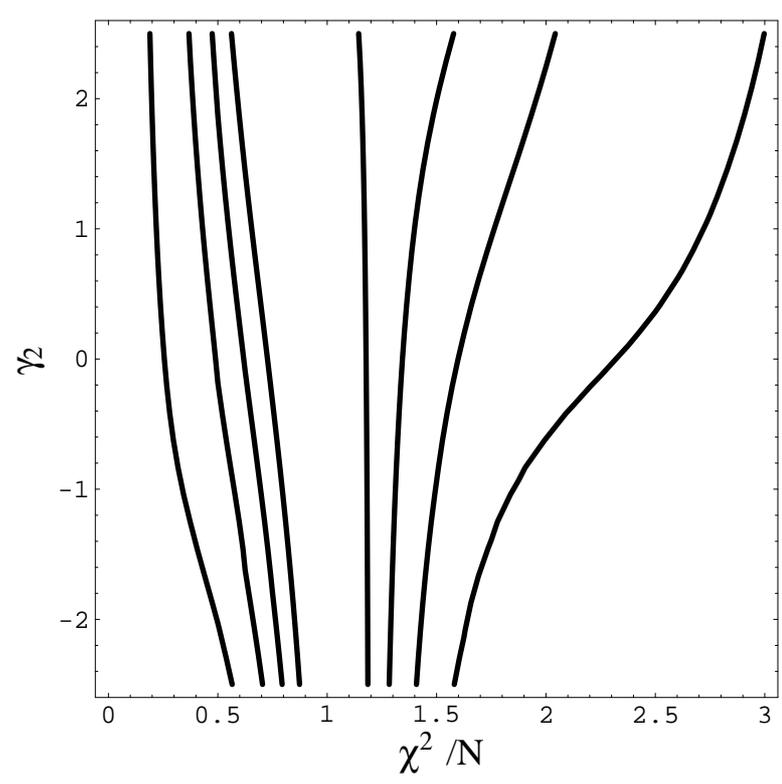
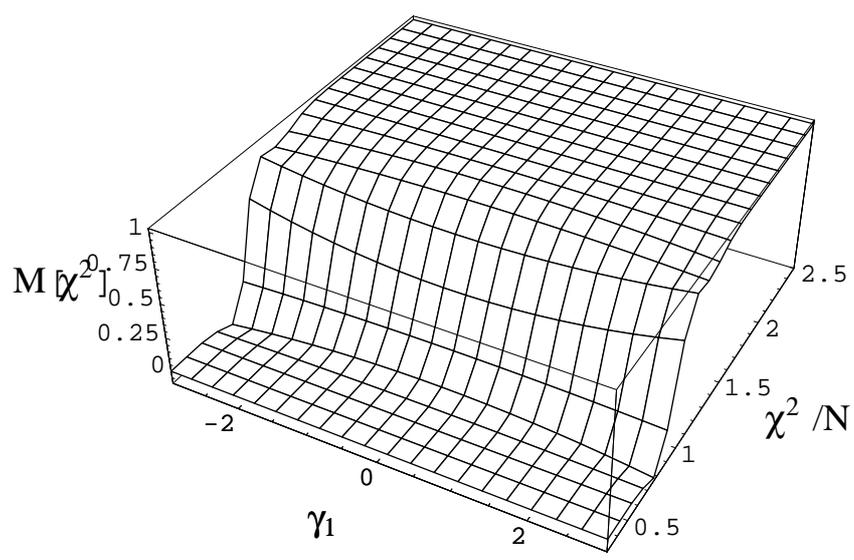
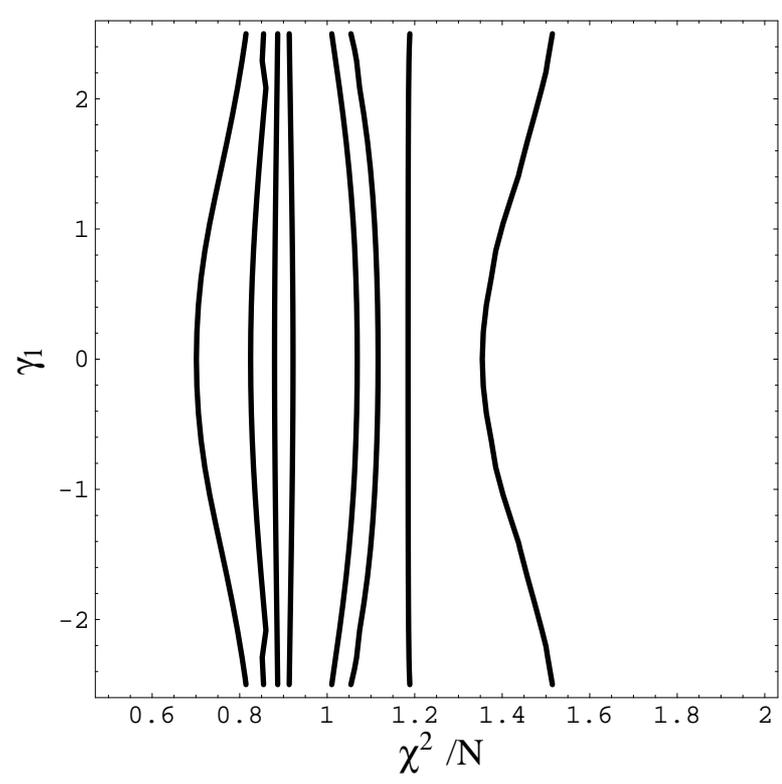
\end{document}